\documentclass[twocolumn,showpacs,preprintnumbers,amsmath,amssymb]{revtex4}

\usepackage{graphicx}
\usepackage{dcolumn}
\usepackage{bm}

\newcommand{\mean}[1]{\langle{#1}\rangle}
\newcommand{\pro}[2]{\langle{#1}|{#2}\rangle}
\newcommand{\bra}[1]{\langle{#1}|}
\newcommand{\ket}[1]{|{#1}\rangle}

\newcommand{\Tr}{{\rm Tr}\hspace{0.07cm}}

\newcommand{\cPr}[1]{{\rm Pr}\left({#1} \right)}

\begin{document}

\title{Information amplification via postselection: 
A parameter estimation perspective
}

\author{Saki Tanaka}%
\email[E-mail address: ]{saki-tanaka@a6.keio.jp}
\author{Naoki Yamamoto}
\email[E-mail address: ]{yamamoto@appi.keio.ac.jp}
\affiliation{Department of Applied Physics and Physico-Informatics, 
Keio University, Yokohama 223-8522, Japan}
\date{\today}%

\begin{abstract}

It is known that weak measurement can significantly amplify the mean of measurement results, sometimes out
of the range limited in usual quantum measurement. This fact, as actively demonstrated recently in both theory
and experiment, implies the possibility to estimate a very small parameter using the weak measurement technique.
But does the weak measurement really bring about the increase of ginformationh for parameter estimation? This
paper clarifies that, in a general situation, the answer is NO; more precisely, the weak measurement cannot further
decrease the lower bound of the estimation error,
i.e. the so-called CramLer-Rao bound, which is proportional to
the inverse of the quantum Fisher information.

\end{abstract}


\maketitle


\section{Signal amplification and parameter estimation}

The importance of quantum metrology is self-evident in
a wide area of applications such as the atomic clock and
gravitational wave detection \cite{Giovannetti}.
The most simple form of
this problem is to estimate an unknown small parameter
$\theta$ contained in the unitary evolution
$\hat{U}=\exp(-i\theta \hat{H})$, where the Hamiltonian 
$\hat{H}$ is assumed to be known.

Among various approaches to this problem, a specifically attractive 
one is the method based on {\it weak measurement} \cite{AAVPRL1988}, 
in the situation where $\hat{H}$ is an interaction Hamiltonian and 
we want to estimate the interaction strength $\theta$.
The method is briefly described as follows; 
For a system ${\cal H}$ and a probe ${\cal K}$, an interaction 
Hamiltonian $\hat{H}=\hat{A}^{\cal H}\otimes \hat{p}^{\cal K}$ with 
$\hat{p}^{\cal K}$ the probe momentum operator is given to us. 
(In what follows we will omit the subscript ${\cal H}$ or ${\cal K}$ 
when obvious.) 
Also, we are allowed to freely set a system's initial state 
$\ket{i}_{\cal H}$ and a final state $\ket{f}_{\cal H}$, which are 
respectively called pre and post selection. 
Then, for a small $\theta$, the probe position operator 
$\hat{x}^\mathcal{K}$ satisfying $[\hat{x}, \hat{p}]=i$ experiences 
a shift proportional to the {\it weak value} 
$\mean{ \hat{A} }_w := \mean{f|\hat{A}|i}/ \mean{f|i}$; 
in fact, the mean value is given by (see \cite{CWV}) 
\begin{align}
    \mean{ \hat{x}^\mathcal{K} } 
      \propto \theta \cdot \mathrm{Re} \mean{ \hat{A} }_w. 
\end{align}
This implies that, by choosing a nearly orthogonal pair of 
$\ket{i}$ and $\ket{f}$, we obtain a largely amplified signal 
$\mean{\hat{x}^\mathcal{K}}$, which would give us a chance to 
estimate $\theta$ highly accurately. 
This signal's amplification technique was originally developed by 
Hosten {\it et al.} in an application to detect the spin hall effect of 
light \cite{Hosten}. 
Also Dixson {\it et al.} have demonstrated the detection of a slight 
tilt of a mirror in a Sagnac interferometer \cite{Dixson}. 
Furthermore, in some recent works \cite{WuLi,KT,Nakamura,Nishizawa,
Susa2012,LeeTsutsui} it was clarified that the amplification 
is still possible to a certain extent even when $\theta$ is 
not small.

The above-described method, however, lacks a statistical viewpoint 
for analyzing how accurate we can estimate the parameter $\theta$. 
In other words, rather than the mean, we should evaluate the 
estimation error, based on quantum statistics 
\cite{Helstrom,Holevo,Braunstein,Holevo2001,Hayashi}. 
Especially in our case we invoke the theory of a one-parameter 
estimation described as follows; 
When $n$ independent copies of a state $\hat{S}_\theta$ with single 
parameter $\theta$ are given to us, any estimator (an observable to 
be measured) $\hat{T}$ is limited in estimation performance 
by the {\it quantum Cram\'{e}r-Rao inequality}
\begin{align}
\label{eq:QCRineq}
    \mean{ (\hat{T} - \theta \hat{I})^2 }_{\hat{S}_\theta}
       ={\rm Tr}\big[ \hat{S}_\theta (\hat{T} - \theta \hat{I})^2 \big]
          \ge \frac{1}{n I(\hat{S}_\theta)}. 
\end{align}
Here, $I(\hat{S}_\theta)$ is the {\it SLD quantum Fisher information}: 
\begin{align}
\label{eq:defSLD-Fihser}
    I(\hat{S}_\theta) 
       := \langle \hat{L}_\theta^2 \rangle_{\hat{S}_\theta}
        = {\rm Tr}\big[\hat{L}_\theta^2 \hat{S}_\theta \big], 
\end{align}
where $\hat{L}_\theta$ is a Hermitian operator called the 
{\it symmetric logarithmic derivative (SLD)} satisfying the 
following linear algebraic equation:
\begin{align}
\label{eq:defSLD}
    \partial_{\theta}\hat{S}_\theta 
      = \frac{1}{2} \left( 
          \hat{S}_\theta \hat{L}_\theta 
             + \hat{L}_\theta \hat{S}_\theta \right). 
\end{align}
In this paper, we simply call Eq.~\eqref{eq:defSLD-Fihser} the Fisher 
information. 
Equation \eqref{eq:QCRineq} means that a state with larger Fisher 
information allows us to estimate the parameter $\theta$ with better 
accuracy. 
Actually, despite that the Cram\'{e}r-Rao bound $1/n I(\hat{S}_\theta)$ 
generally depends on the unknown parameter $\theta$, there have been 
developed some estimation techniques to attain the equality in 
Eq.~\eqref{eq:QCRineq} \cite{Nagaoka,Fujiwara_MLE}.

From the above discussion, we should turn our attention from the 
mean to the Fisher information, for evaluating possible advantages 
of the weak measurement technique in signal amplification in the 
sense of parameter estimation. 
That is, our question is the following; 
{\it Does the weak measurement amplify the information for 
parameter estimation, in the sense of Fisher information multiplied 
by the number of copies of the state (i.e. the inverse of the 
Cram\'{e}r-Rao bound)?} 
Actually this problem has been studied by Knee {\it et al}. in \cite{Knee} 
in a specific example and they found that the Cram\'{e}r-Rao bound 
cannot be decreased by weak measurement (and more broadly by 
postselection as mentioned later). 
This fact leads us to have a negative impression for the use of 
weak measurement technique in parameter estimation problems. 
The main contribution of this paper is to clarify that the answer 
to the above question is NO; 
that is, we prove that, in a general situation, the weak measurement 
cannot decrease the Cram\'{e}r-Rao bound.

Before closing this section, we make two remarks. 
First, in the literature there are several works addressing 
the weak measurement in the framework of parameter estimation, 
particularly with {\it classical} Fisher information 
\cite{Hofmann,Shikano,Strubi,Viza}. 
Second, as motivated by the previous results \cite{WuLi,KT,Nakamura,
Nishizawa,Susa2012,LeeTsutsui}, we will work on the subject without 
assuming that $\theta$ is small. 
In this sense, the scheme is not anymore what is based on weak 
measurement, rather at the heart of the scheme is the postselection; 
hence the above question is a bit modified.
Note that this problem setting also discerns our work and \cite{Knee} 
from \cite{Hofmann,Shikano,Strubi,Viza}.


\section{Quantum Fisher information of the postselected state}

\begin{figure}[t!]
\centering 
\includegraphics[width=7.7cm]{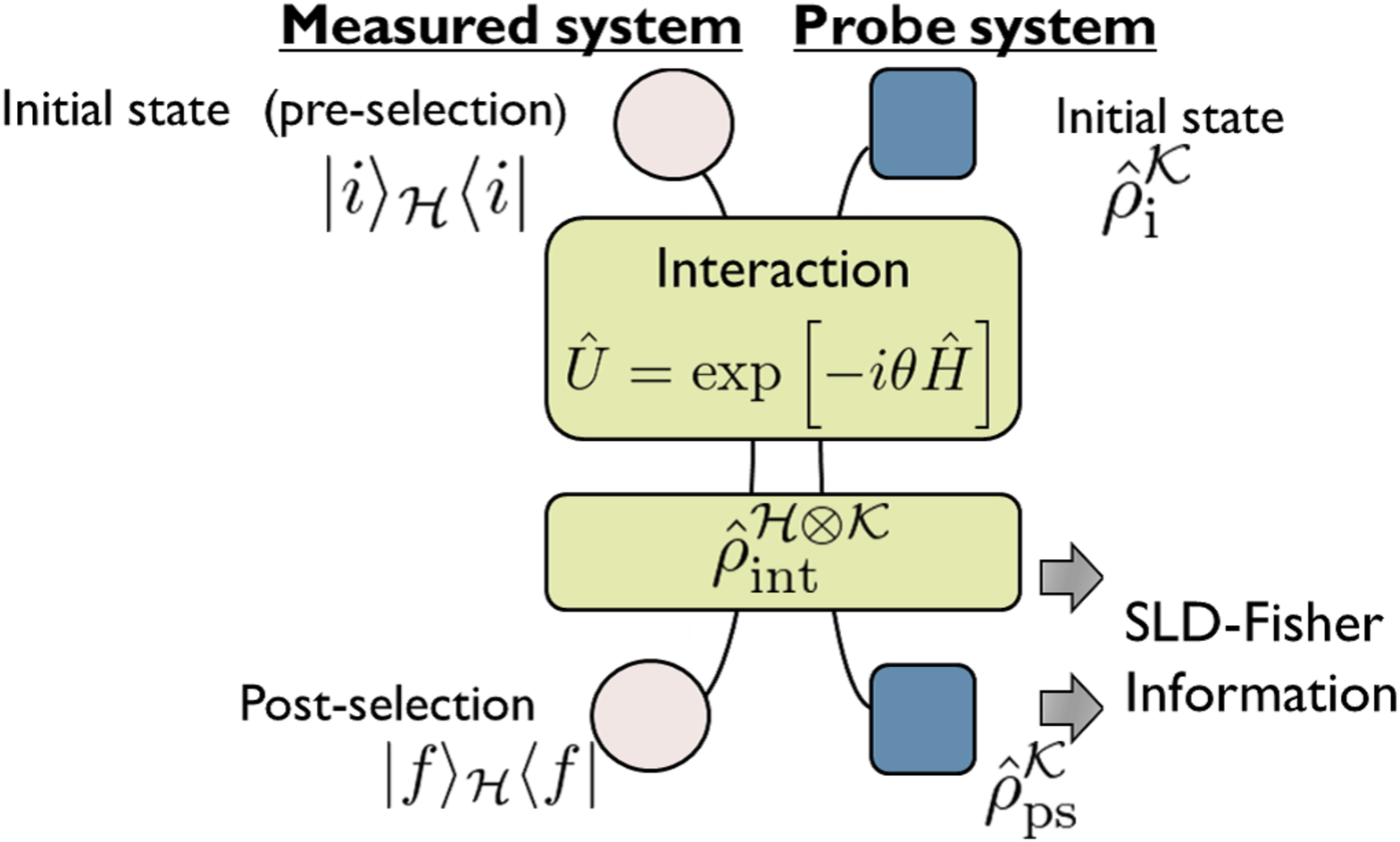}
\caption{
Procedure of the postselection.}
\label{fig:model}
\end{figure}

We study the composition of a system ${\cal H}$ and a probe ${\cal K}$ 
with initial state 
$\ket{i}_{\cal H}\bra{i}\otimes\hat{\rho}_\mathrm{i}^\mathcal{K}$. 
This state is subjected to the interaction 
$\hat{U}=\exp(-i\theta \hat{H})$, which yields 
\begin{equation}
\label{whole state}
    \hat{\rho}_\mathrm{int}^{\mathcal{H}\otimes \mathcal{K}}
       =\hat{U} (\ket{i}_\mathcal{H}\bra{i} 
              \otimes \hat{\rho}_{\rm i}^\mathcal{K}) \hat{U}^\dagger. 
\end{equation}
Again, $\theta$ is a single unknown parameter. 
We then consider a specific state with its system component 
projected onto a fixed state $\ket{f}_{\cal H}$, i.e., the 
following {\it postselected} (ps) state (see Fig.~\ref{fig:model}): 
\begin{equation}
\label{ps state}
   \hat{\rho}^\mathcal{K}_\mathrm{ps} 
    = \frac{ {\rm Tr}_{\mathcal{H}}\big[ 
            ( \ket{f}_\mathcal{H}\bra{f} \otimes \hat{I}^\mathcal{K}) 
                \hat{\rho}_\mathrm{int}^{\mathcal{H} \otimes \mathcal{K}} ]}
              {\Pr(f)}
    = \frac{ \hat{B} \hat{\rho}_\mathrm{i}^\mathcal{K} \hat{B }^\dagger }
       {{\rm Tr} \big(
         \hat{B}\hat{\rho}_\mathrm{i}^\mathcal{K} \hat{B}^\dagger \big)}, 
\end{equation}
where 
$\hat{B} = \mbox{}_{\cal H}\bra{f} \hat{U} \ket{i}_\mathcal{H}$ 
and $\Pr(f)$ is the success probability of the postselection: 
\begin{align}
\label{success prob}
    {\rm Pr}(f)
      = {\rm Tr}\big[ 
            (\ket{f}_\mathcal{H}\bra{f} \otimes \hat{I}^\mathcal{K}) 
              \hat{\rho}_{\rm int}^{ \mathcal{H} \otimes \mathcal{K} } \big]
      &= {\rm Tr} \big(
          \hat{B}\hat{\rho}_\mathrm{i}^\mathcal{K} \hat{B}^\dagger \big). 
\end{align}

What we are concerned with is, under the assumption that both 
${\cal H}$ and ${\cal K}$ are accessible as in the case of 
\cite{Hosten,Dixson}, if the postselected state \eqref{ps state} 
would contain more valuable information than the whole state 
without conditioning, \eqref{whole state}. 
Hence here we can formulate our first problem; 
{\it is the Fisher information of the state \eqref{ps state} bigger 
than that of the state \eqref{whole state}?} 
In general, it is not straightforward to calculate the Fisher 
information, but in the case of pure states it is uniquely and 
explicitly obtained. 
That is, for a pure state 
$\hat{S}_\theta = \ket{\chi_\theta}\bra{\chi_\theta}$, the Fisher 
information is given by 
\[
    I(\hat S_\theta)
    = \langle \hat{L}^2 \rangle_{\hat{S}_\theta} 
     = 4 \left( 
       \mean{ \partial_\theta  \chi_\theta | \partial_\theta \chi_\theta} 
     - \bigr| \mean{ \partial_\theta \chi_\theta | \chi_\theta } \bigr|^2
         \right), 
\]
where $\ket{\partial_\theta \chi_\theta}
=\partial \ket{\chi_\theta}/\partial\theta$. 
To use this formula, let us assume 
$\rho_{\rm i}^{\cal K}=\ket{\psi}_{\cal K}\bra{\psi}$. 
Then, the Fisher information of the state \eqref{whole state} is 
obtained as 
\begin{align}
    I( \hat{\rho}^{\mathcal{H} \otimes \mathcal{K}}_\mathrm{int} )
      & = 4  \left[ \bra{i, \psi} \hat{H}^2 \ket{i, \psi} 
                    - \bra{i, \psi}\hat{H}\ket{i,\psi}^2 \right]
\notag \\
      & = 4 \bigl( \mean{\hat{H}^2} - \mean{\hat{H}}^2 \bigr), 
\end{align}
while that of the postselected state \eqref{ps state} is given by 
\begin{equation}
\label{eq:SLDPPSProb}
    I(\hat{\rho}^\mathcal{K}_\mathrm{ps})
     = \frac{ 4 \langle \psi | \partial_\theta \hat{B}^\dagger 
                       \partial_\theta \hat{B} | \psi \rangle }
            {\Pr(f)}
     - \frac{ 4 |\langle \psi | \partial_\theta \hat{B}^\dagger \hat{B} 
                                 | \psi \rangle |^2 }
            {\Pr(f)^2}. 
\end{equation}
Here the success probability is 
$\Pr(f) = \bra{\psi}\hat{B}^\dagger \hat{B}\ket{\psi}$. 
Equation \eqref{eq:SLDPPSProb} shows that 
$I(\hat{\rho}^\mathcal{K}_\mathrm{ps})$ takes a large number, or 
equivalently the postselected state becomes more valuable, if 
$\Pr(f)$ is taken to be small by the postselection. 
In particular, when $\hat{H} =\hat{A}\otimes \hat{p}$, we find 
that 
\begin{align}
\label{relation to weak value}
    \lim_{\theta \rightarrow 0} 
      I( \hat{\rho}^\mathcal{K}_\mathrm{ps})
       = \big|  \mean{\hat{A}}_w \big|^2 
 \bra{\psi} (\hat{p} - \mean{\hat{p}})^2 \ket{\psi}. 
\end{align} 
Hence, in the weak interaction limit $\theta\rightarrow 0$, 
increasing the weak value $|\mean{\hat{A}}_w|$ via the postselection 
directly means increase of the Fisher information. 
This fact leads us to expect that the signal amplification 
technique based on the weak measurement 
\cite{Hosten,Dixson,WuLi,KT,Nakamura,Nishizawa,Susa2012,LeeTsutsui} 
would be statistically consistent.


\section{Sensitivity amplification via postselection}
\label{sec2states}

In this section, we compare the two Fisher informations presented 
above in a specific example. 
Note that, as seen in Eq.~\eqref{eq:QCRineq}, the Fisher 
information itself does not provide the lower bound of the 
estimation error in a repeated experiment; 
we will discuss this in the next section. 
Here the Fisher information is identified with the 
``distinguishability" of states \cite{Helstrom,Holevo,Braunstein,
Holevo2001,Hayashi}. 
That is, in terms of the Bures distance between two states 
$\hat{S}_1$ and $\hat{S}_2$: 
\[
    b(\hat{S}_1, \hat{S}_2)
     = \Big[ 2 
         - 2\Tr\sqrt{\hat{S}_1^{1/2}\hat{S}_2\hat{S}_1^{1/2}} \Big]^{1/2}, 
\]
the Fisher information gives a metric measuring the small 
shift of a parameter-dependent state $\hat{S}_\theta$ as follows: 
\[
    b(\hat{S}_\theta, \hat{S}_{\theta+d\theta})
     = I(\hat{S}_\theta)(d\theta)^2/4. 
\]
This means that, if the Fisher information takes a large number, 
the state $\hat{S}_{\theta+d\theta}$ is very sensitive to the 
parameter change $d\theta$ and thus can be easily distinguished 
from $\hat{S}_\theta$.

Here we study the following example. 
The system is ${\cal H}={\mathbb C}^2$ and the probe is a one 
dimensional meter device. 
The interaction is given by 
$\exp(-i \theta \hat{H})
=\exp(-i \theta \hat{\sigma}_z^\mathcal{H}\otimes \hat{p}^\mathcal{K})$, 
where $\hat{\sigma}_z^\mathcal{H}$ is the Pauli matrix and 
$\hat{p}^\mathcal{K}$ is the momentum operator. 
Also let the pre and post selected states of $\mathcal{H}$ be 
$\ket{i} = \cos t_1 \ket{0} + e^{i s_1} \sin t_1 \ket{1}$ and 
$\ket{f} = \cos t_2 \ket{0} + e^{i s_2} \sin t_2 \ket{1}$. 
The initial probe state is Gaussian with wave function 
$\pro{p}{\psi} = (2 \sigma^2/\pi)^{1/4} \exp(-\sigma^2 p^2)$. 
Then, the Fisher information of 
$\hat{\rho}_\mathrm{ps}^\mathcal{K}$ is calculated as 
\begin{align}
\label{eq:Fishertwoinfty} 
     I ( \hat{\rho}_\mathrm{ps}^\mathcal{K}) 
       &= 
         \frac{1}{\sigma^2} 
     \frac{
          w_+^2 + \frac{\theta^2}{\sigma^2} 
                       w_+ w_- e^{-\frac{\theta^2}{2 \sigma^2 }} 
                - w_-^2 e^{-\frac{\theta^2}{\sigma^2}}
        }{
         \big( w_+ +w_- e^{-\frac{\theta ^2}{2 \sigma ^2 }} \big)^2
        },
\\
     w_\pm &= \bigl( | \langle f | i \rangle |^2 
              \pm | \langle f | \hat{\sigma}_z | i \rangle |^2 \bigr) 
                {\big /} 2, 
\notag 
\end{align}
while that of 
$\hat{\rho}_\mathrm{int}^{\mathcal{H} \otimes \mathcal{K}}$ is 
given by 
\begin{align}
\label{Fisher int}
    I(\hat{\rho}_\mathrm{int}^{\mathcal{H} \otimes \mathcal{K}}) 
      = 1 \big{/} \sigma^2. 
\end{align}

\begin{figure}[t]
\centering 
\includegraphics[width=5.4cm]{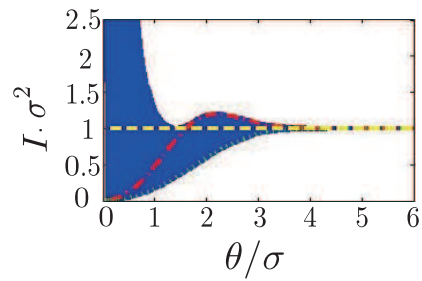}
\caption{
Fisher informations versus the parameter $\theta$. 
The blue region represents the set of curves of 
$I(\hat{\rho}_\mathrm{ps} ^\mathcal{K})$, while the yellow dashed 
line indicates 
$I(\hat{\rho}_\mathrm{int}^{\mathcal{H}\otimes \mathcal{K}})$. 
The red dashed-dotted and green dotted lines show the classical Fisher 
information $I_\mathrm{c}(f_{\rm ps}(x))$ with $c=1$ and $c=-1$, 
respectively. 
}
\label{fig:SLDPPS}
\end{figure}

Figure \ref{fig:SLDPPS} shows the Fisher informations 
\eqref{eq:Fishertwoinfty} and \eqref{Fisher int} versus the 
parameter $\theta$. 
The blue region represents the set of curves of 
$I(\hat{\rho}_\mathrm{ps}^\mathcal{K})$ generated with various 
values of the parameters $(t_1, t_2, s_1-s_2)$, while the yellow 
dashed line indicates 
$I(\hat{\rho}_\mathrm{int}^{\mathcal{H}\otimes \mathcal{K}})$. 
This figure shows that, in a certain range of $\theta$, the appropriate 
postselection brings about the increase of Fisher information; 
especially for small $\theta$, $I(\hat{\rho}^\mathcal{K}_\mathrm{ps})$ 
becomes infinitely large when $\ket{f}$ is nearly orthogonal 
to $\ket{i}$, which is indeed expected from 
Eq.~\eqref{relation to weak value}. 
As a summary, the postselected state can become more sensitive to 
the parameter change and thus, in this sense, contain more valuable 
information than the whole state without conditioning.

Here note that the quantum Fisher information does not depend 
on how we actually measure the system and extract information from it. 
Because of this fact, the quantum Fisher information is always bigger 
than any classical Fisher information of a probabilistic distribution 
generated from a certain fixed measurement. 
Thus, to maintain practical superiority of the postselection, 
we need to show that the classical Fisher information associated 
with $\hat{\rho}_\mathrm{ps}^\mathcal{K}$ is bigger than that of 
$\hat{\rho}_\mathrm{int}^{\mathcal{H} \otimes \mathcal{K}}$. 
Specifically here let us consider measuring the probe position 
operator $\hat{x}^\mathcal{K}$. 
Then, when $\cos^2 t_1 =\cos^2 t_2 =1/2$ and $c:= \cos(s_1-s_2) = \pm 1$, 
the probabilistic distribution is calculated as 
\[
    f_\mathrm{ps}(x) 
       =\frac{
          e^{-\frac{(x- \theta)^2 }{2 \sigma^2} }
          + e^{-\frac{(x+ \theta)^2 }{2 \sigma^2} }
          + 2ce^{-\frac{x^2 + \theta^2 }{2 \sigma^2}}  }
             {
          2 \sqrt{2\pi \sigma^2 } 
             \big( 1 + c e^{-\frac{\theta^2 }{2 \sigma^2}} \big)}, 
\]
and its classical Fisher information is 
\begin{align}
   I_\mathrm{c} \bigl(f_\mathrm{ps}(x) \bigr) 
     & := \int_{-\infty}^\infty dx ~
             \bigl\{ \partial_\theta \log f_\mathrm{ps}(x) \bigr\}^2~
                 f_\mathrm{ps}(x) 
\notag \\
     & = \frac{1}{\sigma^2} 
         \frac{
            1 + c \frac{\theta^2}{\sigma^2} e^{-\frac{\theta^2}{2 \sigma^2}} 
              - e^{-\frac{\theta^2}{\sigma^2}} }
              { \big( 1 + e^{-\frac{\theta^2}{2\sigma^2} } \big)^2}. 
\notag
\end{align}
Figure \ref{fig:SLDPPS} shows that 
$I_\mathrm{c} \bigl(f_\mathrm{ps}(x) \bigr)$ with $c=1$ reaches the 
quantum Fisher information $I(\hat{\rho}_\mathrm{ps}^\mathcal{K})$ 
in the range where 
$I(\hat{\rho}_\mathrm{ps}^\mathcal{K}) \geq 
I(\hat{\rho}_\mathrm{int}^{\mathcal{H}\otimes \mathcal{K}})$ holds. 
Clearly, in this case, $I_\mathrm{c} \bigl(f_\mathrm{ps}(x) \bigr)$ 
is bigger than any classical Fisher information associated with 
$\hat{\rho}_\mathrm{int}^{\mathcal{H}\otimes \mathcal{K}}$. 
Thus, by measuring $\hat{x}^\mathcal{K}$ with $c=1$, we can 
indeed extract more information by the postselection.

\section{Comparing estimation errors in asymptotic condition}

In the last section, we have seen that the sensitivity of the 
state to the parameter $\theta$ can be enhanced via the 
postselection. 
This would suggest, as shown in Eq.~\eqref{eq:QCRineq}, that the 
postselection can bring about further decrease of the Cram\'{e}r-Rao 
bound, i.e, the strict lower bound of the estimation error of $\theta$. 
However, the critical issue with this postselection technique 
is that we obtain the state only when the postselection succeeds; 
that is, we have to construct the estimator using less measurement 
data, compared to the standard method based on the whole state 
$\hat{\rho}^{\mathcal{H} \otimes \mathcal{K}}_\mathrm{int}$. 
Therefore, the problem becomes comparing the Cram\'{e}r-Rao bounds 
$1/n_{\rm ps}I(\hat{\rho}_\mathrm{ps}^\mathcal{K})$ and 
$1/n_{\rm int}I(\hat{\rho}^{\mathcal{H} \otimes \mathcal{K}}_\mathrm{int})$, 
where $n_{\rm ps}$ and $n_{\rm int}$ are the number of trials in 
those methods, respectively.

The above problem is not straightforward to solve. 
However, if we are allowed to perform the trial infinitely 
many times, it is possible to have a general answer. 
In fact, in such asymptotic condition, there exist estimators 
attaining the Cram\'{e}r-Rao bounds \cite{Nagaoka,Fujiwara_MLE}, 
and furthermore, the number of trials are explicitly given by 
$n_{\rm ps}={\rm Pr}(f)n$ and $n_{\rm int}=n$, as was also 
discussed in \cite{Knee}. 
Hence, the problem is now to compare 
${\rm Pr}(f)I(\hat{\rho}_\mathrm{ps}^\mathcal{K})$ and 
$I(\hat{\rho}^{\mathcal{H} \otimes \mathcal{K}}_\mathrm{int})$.

To solve the problem, let us define 
\begin{align}
   \tilde{B} := \mbox{}_{\cal H}\bra{f} 
                  e^{-i\theta (\hat{H}- \mean{\hat{H}})} 
                     \ket{i}_\mathcal{H}, 
\end{align}
where $\mean{\hat{H}} 
= \Tr(\hat{\rho}_\mathrm{int}^{\mathcal{H} \otimes \mathcal{K}} \hat{H})$, 
which leads to 
\[
    \hat{\rho}^\mathcal{K}_\mathrm{ps} = 
      \tilde{B} \hat{\rho}_i^\mathcal{K} \tilde{B}^\dagger / 
           {\rm Tr}(\tilde{B} \hat{\rho}_i^\mathcal{K} \tilde{B}^\dagger),~~
    \cPr{f} = \bra{\psi} \tilde{B}^\dagger \tilde{B} \ket{\psi}. 
\]
Then, since $\ket{f}_\mathcal{H}\bra{f}\otimes I^{\mathcal{K}} \leq 
\hat{I}^{\mathcal{H} \otimes \mathcal{K}}$, we have 
\begin{align} 
    & \hspace{-1em} 
    \cPr{f} I ( \hat{\rho} _\mathrm{ps} ^{ \mathcal{K} } ) /4
\notag \\
    & = 
     \bra{\psi} \partial_\theta \tilde{B}^\dagger \partial_\theta \tilde{B}
          \ket{\psi}
     - \frac{\bigl| \bra{\psi} \partial_\theta \tilde{B}^\dagger \tilde{B}
                \ket{\psi} \bigr| ^2} 
            {\bra{\psi} \tilde{B}^\dagger \tilde{B} \ket{\psi}}
\notag \\
    & \leq 
     \bra{\psi} \partial_\theta \tilde{B}^\dagger \partial_\theta \tilde{B}
          \ket{\psi}
\notag \\
    & = \mbox{}_\mathcal{K}\bra{\psi} \mbox{}_\mathcal{H} \bra{i}
          (\hat{H} - \mean{\hat{H}})
             e^{-i \theta(\hat{H}- \mean{\hat{H}})} \ket{f}_\mathcal{H}
\notag \\
    &~~~~~~~ \times 
      \mbox{}_\mathcal{H}\bra{f}
         (\hat{H}- \mean{\hat{H}})
             e^{i \theta (\hat{H}- \mean{\hat{H}})} 
      \ket{i}_\mathcal{H} \ket{\psi}_\mathcal{K}
\notag \\
    & \leq \bra{i, \psi} (\hat{H}- \mean{\hat{H}})^2
           e^{-i \theta(\hat{H}- \mean{\hat{H}}) }  
           e^{i \theta (\hat{H}- \mean{\hat{H}}) } \ket{i, \psi}
\notag \\
    & = \bra{i, \psi} (\hat{H}- \mean{\hat{H}})^2 \ket{i, \psi}
     = I ( \hat{\rho}_\mathrm{int}^{\mathcal{H} \otimes \mathcal{K}})/4. 
\notag
\end{align}
As a result, 
\begin{align}
\label{ieq:psleint}
    \cPr{f} I \bigl( \hat{\rho}_\mathrm{ps}^\mathcal{K} \bigr) 
      \leq 
        I\bigl(\hat{\rho}_\mathrm{int}^{\mathcal{H} \otimes 
                               \mathcal{K}} \bigr). 
\end{align}
Therefore, we now obtain a general answer to the question 
considered throughout the paper; 
in the asymptotic condition, the postselection method 
does never yield a better estimator that outperforms the 
standard method that allows us to perform any measurement on 
the whole composite system. 
Within the context of quantum metrology mentioned at the beginning 
of Sec.~I, this fact has the following interpretation. 
Again, the problem is to estimate the unknown parameter $\theta$ 
contained in the given system ${\cal S}$ in the form 
$\hat{U}=\exp(-i\theta \hat{H})$. 
Then, the weak-value amplification techniques found in the 
literature suggest us to divide the system into two parts 
${\cal H}$ and ${\cal K}$, perform a suitable postselection 
on ${\cal H}$, and then detect a rare event on ${\cal K}$, which 
would contain valuable information about $\theta$. 
However, the inequality \eqref{ieq:psleint} implies that this 
strategy does not have an advantage in estimating $\theta$ for 
{\it any} partitioning of the given system ${\cal S}$ into two 
subsystems, ${\cal S}={\cal H}\otimes{\cal K}$, and for {\it any} 
type of postselection on ${\cal H}$. 
In this sense, the inequality \eqref{ieq:psleint} is a no-go 
theorem in the field of quantum metrology.


\section{Discussion}

\begin{figure}[t]
\centering 
\includegraphics[width=8.6cm]{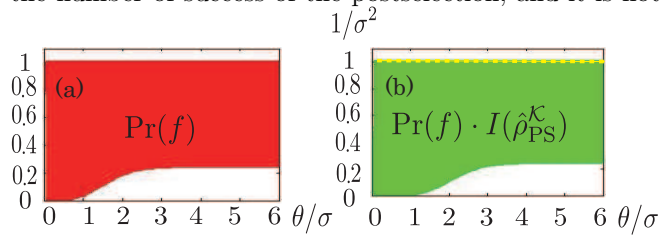}
\caption{
(a) The success probability ${\rm Pr}(f)$ and (b) Fisher 
information normalized by ${\rm Pr}(f)$, as functions of 
$\theta/\sigma$. 
The regions represent the set of curves of ${\rm Pr}(f)$ and 
$\Pr(f) I(\hat{\rho}_\mathrm{ps}^\mathcal{K})$ generated with 
various values of $(t_1, t_2, s_1-s_2)$. 
}
\label{fig:PrfPrfSLD}
\end{figure}

The main conclusion we have obtained is that, in general, the signal 
amplification technique based on the weak measurement or more broadly 
the postselection is useless in the statistics sense. 
Note again that this result is obtained under asymptotic condition; 
in other words, when the measurement can be carried out only finite 
times, ${\rm Pr}(f)n$ does not anymore have the meaning of the number 
of success of the postselection, and it is not clear whether or not 
a similar inequality holds. 
Actually we have the following fact: 
Let us reconsider the example studied in Sec~III. 
Figure~\ref{fig:PrfPrfSLD} shows the success probability 
${\rm Pr}(f)$ and the Fisher information normalized by 
${\rm Pr}(f)$. 
In the right panel, the yellow dashed line shows 
$I(\hat{\rho}_\mathrm{int}^{\mathcal{H} \otimes \mathcal{K}})
=1/\sigma^2$. 
This figure demonstrates that, by performing a suitable postselection, 
it is possible to attain nearly the equality in 
Eq.~\eqref{ieq:psleint} for almost all $\theta$; 
hence, it seems that such a fine postselection could realize 
$n_{\rm ps}I(\hat{\rho}_\mathrm{ps}^\mathcal{K})>
n_{\rm int}I(\hat{\rho}^{\mathcal{H} \otimes \mathcal{K}}_\mathrm{int})$ 
for finite numbers of trial $n_{\rm ps}$ and $n_{\rm int}$. 
However, Ferrie and Combes proved in \cite{Ferrie} that this conjecture 
does not hold; 
that is, the postselection does not enhance the precision of 
the parameter estimate for any amount of data. 
We also should point out the recent preprint \cite{Knee 2013} by 
Knee and Gauger, which proves no advantage of the postselection-based 
amplification technique in a slightly different setting.

Another important question is about how to experimentally 
demonstrate the inequality \eqref{ieq:psleint}. 
To achieve this goal, we need to construct a system such that 
we can measure the whole system {\it globally} for computing 
$I(\hat{\rho}_\mathrm{int}^{\mathcal{H} \otimes \mathcal{K}})$ 
as well as each subsystem {\it locally} for computing 
$I(\hat{\rho}_\mathrm{ps}^\mathcal{K})$. 
For instance a pair of trapped ions, which corresponds to 
$\mathcal{H} \otimes \mathcal{K}$, fulfills these requirements; 
actually, Riebe, {\it et al.} showed in \cite{experiment} that 
it is possible to couple two trapped calcium ions, manipulate 
each ion individually, and perform a complete global (Bell) 
measurement by detecting fluorescence from the ions. 
On the other hand, for instance a nano-mechanical oscillator 
($\mathcal{H}$) driven by optical force with unknown strength 
$\theta$, which arises due to the interaction between the oscillator 
and an environment field ($\mathcal{K}$), is not a suitable system 
for the experimental demonstration, because in this case the 
environment field is not accessible. 
Here we remark that this issue further raises the following important 
questions: 
Can we apply the postselection technique to estimate unknown 
parameters of an {\it open} system? 
If this is the case, does the postselection offer any advantage? 
Parameter estimation problems for an open system now constitute 
an important research area \cite{Mankei}, so the applicability of 
the postselection technique should be explored.


\end{document}